\documentclass[aps,prb,onecolumn,superscriptaddress]{revtex4-1}
\usepackage{amsmath}
\usepackage{graphics}
\usepackage{graphicx}
\usepackage{epsfig}
\usepackage{amssymb}
\usepackage{amsfonts}
\usepackage{color}
\usepackage{lineno}
 
\binoppenalty=\maxdimen
\relpenalty=\maxdimen

\begin{document}

\title{Hidden Granular Superconductivity Above 500K  in off-the-shelf graphite materials}

\author{Rapha\"el \surname{Rousset-Zenou}}\affiliation{Institut N\'eel, Universit\'e Grenoble Alpes and Centre National de la Recherche Scientifique \\25 rue des Martyrs - BP 166, 38042, Grenoble cedex 9 France}

\author{Samar \surname{Layek}}
\altaffiliation[Present Address: ]{Department of Physics, School of Engineering, University of Petroleum and Energy Studies (UPES), Dehradun, Uttarakhand 248007, India}
\affiliation{Institut N\'eel, Universit\'e Grenoble Alpes and Centre National de la Recherche Scientifique \\25 rue des Martyrs - BP 166, 38042, Grenoble cedex 9 France}

\author{Miguel \surname{Monteverde}}\affiliation{Laboratoire de Physique des Solides, CNRS-Universit\'e Paris-Sud UMR 8502, \\91405 Orsay Cedex, France }

\author{Fr\'ed\'eric \surname{Gay}}\affiliation{Institut N\'eel, Universit\'e Grenoble Alpes and Centre National de la Recherche Scientifique \\25 rue des Martyrs - BP 166, 38042, Grenoble cedex 9 France}

\author{Didier \surname{Dufeu}}\affiliation{Institut N\'eel, Universit\'e Grenoble Alpes and Centre National de la Recherche Scientifique \\25 rue des Martyrs - BP 166, 38042, Grenoble cedex 9 France}

\author{Manuel  \surname{N\'u\~nez-Regueiro}}
\email{Corresponding author. E-mail: manolo.nunez-regueiro@neel.cnrs.fr }
\affiliation{Institut N\'eel, Universit\'e Grenoble Alpes and Centre National de la Recherche Scientifique \\25 rue des Martyrs - BP 166, 38042, Grenoble cedex 9 France}

\date{\today }

\begin{abstract}

It has been reported that graphite hosts room temperature superconductivity.  Here we provide new results that confirm these claims on different samples of highly oriented pyrolytic graphite (HOPG) and commercial flexible graphite gaskets (FGG). After subtraction of the intrinsic graphite diamagnetism, magnetization measurements show convoluted ferromagnetism and superconducting-like hysteresis loops. The ferromagnetism is deconvoluted by fitting with a sigmoidal function and subtracting it from the data. The obtained superconducting-like hysteresis loops are followed to the highest available temperature, 400K. The extrapolation of the decrease of its moment width with temperature indicates a transition temperature  T$_{c}$$\sim$ 550K$\pm$50K for all samples. Electrical resistance measurements confirm the existence at these temperatures of a transition in HOPG samples, albeit without percolation. Besides, the FGG show transitions at temperatures (70K, 270K) near to those reported previously on intercalated-deintercalated graphite, confirming the general character of these superconducting transitions. These results are the first steps in the unveiling of the above room temperature superconductivity of graphite. 
\end{abstract}
\maketitle
\onecolumngrid

\textbf{Introduction}

The quest for superconducting materials started with metals like mercury\cite{Onnes} with superconducting critical temperature T$_c$=4.1K more than a century ago. Then alloys from the A-15 compounds, reached   a T$_c$=22.3K in Nb$_3$Ge\cite{NbGe3}. Followed oxides\cite{Bednorz}, that allowed to reach liquid nitrogen temperatures with YBCO (T$_c$=90K), the highest value for cuprates being T$_c$=166K in fluorinated Hg-1223, under a pressure of 26GPa\cite{Hg1223F}. Lately, hydrogen derived materials in the extremely high pressures of 200GPa, scratched the room temperature goal  \cite{Eremets,Dias}.

On the other hand, for more than twenty years there have been recurrent claims for the existence of unidentified microscopic amounts of ferromagnetism and room temperature superconductivity in different types of graphite, i.e. highly oriented pyrolytic graphite (HOPG)  and natural single crystals of AB (Bernal)  and ABC (rhombohedral)  \cite{EsquinaziGraf,Water,Persistent}. Its origin has been tracked down to granular superconductivity in graphite interfaces\cite{Heikila}, and its transition temperature is obviously above room temperature.

While, in the last few years, Chern ferromagnetic insulating states\cite{Chen} and associated superconductivity \cite{Huang} have been observed in nano devices made of a  few layers of AA graphene twisted at a magic angle. The generated Moir\'e patterns can be viewed as hyper atoms that yield extremely flat bands\cite{Mayou, MacDonald}. Although the superconducting transition temperatures for these systems are in the range of liquid helium, they stress the need of a deeper inspection of the possible superconducting properties of all type of graphene stacking materials. 

Both studies on layered carbon aim at determining the ferromagnetic and superconducting properties of pure graphite, the former being a top-bottom while the latter a bottom-up approach. Whereas the second implies extremely difficult to make but very controlled and defined samples, the first is more like finding a needle in a haystack.  For example, if we believe that "magic angles" can explain the claims of superconductivity in bulk graphite, due to the innumerable angles with which graphene planes can be stacked in a macroscopic sample, the amount of regions having the appropriate geometry should be really tiny and difficult to pin down.  

Recently, an independent confirmation of room temperature superconductivity in graphite has been reported\cite{Layek}. Superconducting transitions with several high temperature T$_c$'s,110K, 245K, 320K, have been observed in KC$_8$ deintercalated at room temperature\cite{Layek}(IDI samples). The low kinetics of such transformation induces a bulk twisted graphite that scans different twist angles and dopings, yielding a small number of crystallites that have the properties for being superconductors. The superconductivity is granular, although almost percolation has been found for the transition at 245K. 

Tracking down with a magnetometer the T$_c$ of a granular superconductivity, due to a few isolated superconducting grains of nanoscopic size in a diamagnetic material such as graphite, is far from evident. The first obvious step is to subtract the large linear intrinsic diamagnetic contribution. Often we are left with a mixed ferromagnetic-superconducting-like hysteresis loop. It must then be deconvoluted fitting a sigmoidal ferromagnetic signal that is then subtracted\cite{Ursula,Layek}, yielding the superconducting superconducting-like hysteresis loop. The variation of this cycle with temperature is then measured. The observed cycle shrinking must be followed\cite{Layek} till its disappearance at T$_c$. The formula derived from Bean's theory\cite{Bean} used in applied superconductivity to estimate the critical current J$_c$ is then employed, i.e. the cycle thickness $\Delta$M(H,T)$\sim$J$_c$(H,T)$\sim$[1-(T/T$_c$)$^ 2$] . Fitting\cite{Layek} the data to this expression then determines T$_c$. In the case where T$_c$ is beyond the temperature range of the magnetometer, its uncertainty will be larger the farther away it is.

We have applied this methodology to probe the existence of above room temperature superconductivity on several samples of commercial, off-the-shelf, forms of graphite. Namely, HOPG and flexible graphite gaskets for industrial applications.

 \textbf{Results}
 
\begin{figure}[ht]
\centering
\includegraphics[width=15cm]{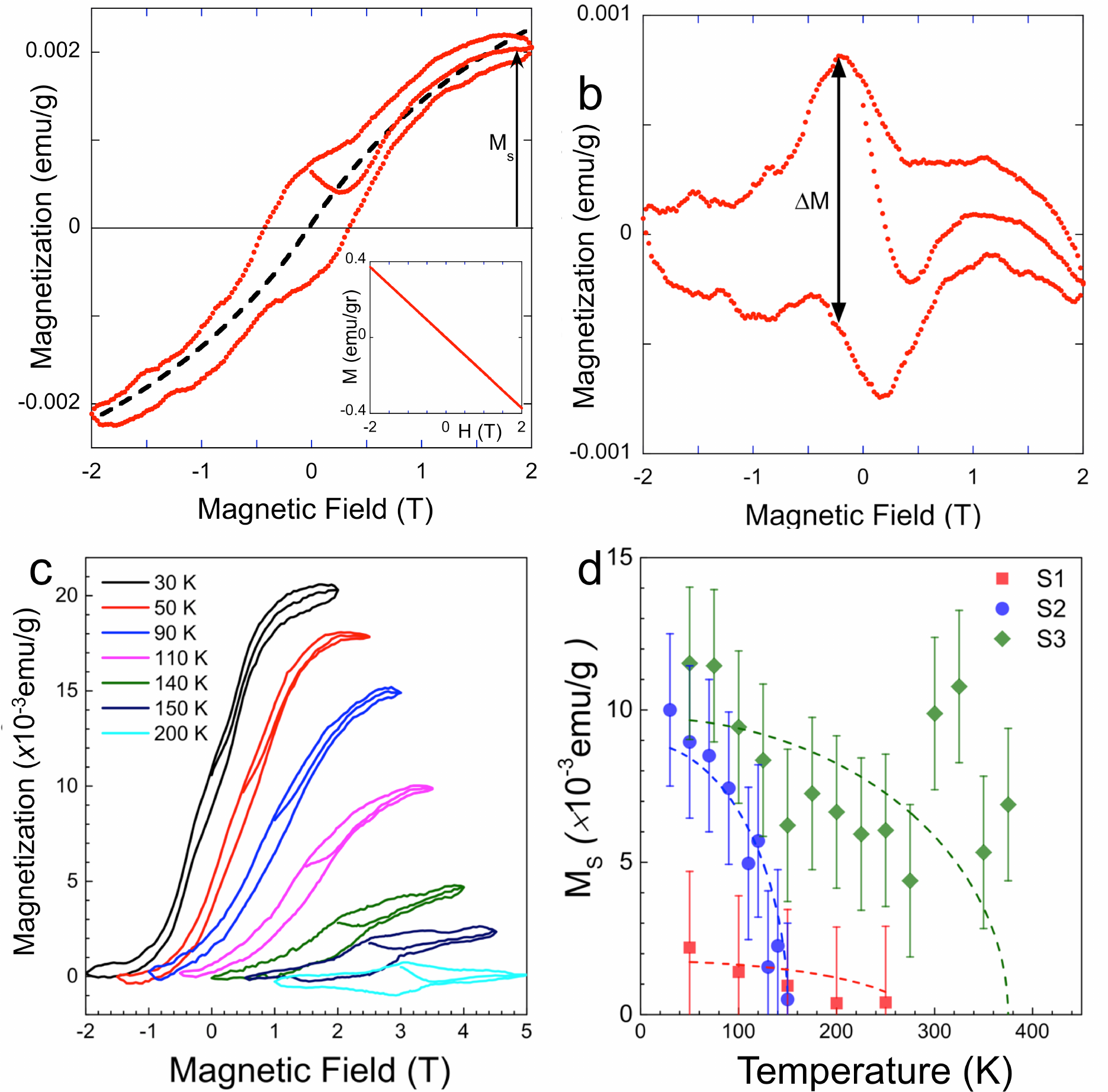}
\caption{ (a)  Ferromagnetic hysteresis cycle after subtraction of the graphite diamagnetic contribution of S1 at 50K. An unexpected bulge due to superconductivity is present around zero (H//c). Dashed line: Ferromagnetic type sigmoidal fit. inset: raw data showing the strong linear diamagnetic magnetization that must be subtracted to obtain the "hidden" ferromagnetic signal. (b) Superconducting-like hysteresis loop after subtraction of the ferromagnetic sigmoidal fit typical of a superconducting state. (c) Ferromagnetic cycles obtained after subtraction of the diamagnetic graphite contribution at different temperatures for S2. For clarity reasons we have defined the zero of magnetization at the beginning of the cycle and shifted each curve by 0.5T. At 150K, ferromagnetism has almost disappeared, and the cycle has become completely diamagnetic at 200K. (d) Saturation magnetizations as a function of temperature for samples S1 (red squares), S2 (blue circles) and S3 (green diamonds). Dashed curves: fits with the phenomenological function (\ref{Ferro}). }
\label{Sigma} 
\end{figure}

We start describing results on HOPG, where room temperature superconductivity was first reported\cite{EsquinaziGraf}. We show results for three samples, S1 and S2 issued from the same HOPG ZYA slab, and S3 from a HOPG ZYH (see Table I for a summary of all samples). 
We show on Fig.~\ref{Sigma}a a superconducting-like hysteresis loop measured at 50K. We see in the inset of Fig.~\ref{Sigma}a the strong linear diamagnetic contribution expected for graphite. The first step in analyzing the data is to subtract the linear in field diamagnetic contribution.  As shown on Fig.~\ref{Sigma}, a ferromagnetic cycle becomes then apparent \cite{EsquinaziGraf,EsquiFerro1}. A clear bulge, though, is present at low fields. If we change the value of the linear diamagnetic correction in $\pm$10\% around the optimal value, the bulge is always visible, even when the ferromagnetic sigmoidal cycle is totally deformed. The shape of the shown curve is similar to that obtained from samples having a mixture of superconductivity and ferromagnetism, e.g. Fig. 4 of Ref.~\onlinecite{Moshschalkov}. 
In these cases the bulge has been demonstrated to correspond to the presence of superconductivity in the sample. Now, we must extract the superconducting from the ferromagnetic signal. For that we use the known procedure to separate the superconducting-like hysteresis loop from the measurement, as described in Refs.~\onlinecite{Layek,Ursula}. A sigmoidal function is fitted, corresponding to the ferromagnetic behavior, and then subtracted  from the data. The deconvoluted superconducting-like hysteresis loop can be observed on  Fig.~\ref{Sigma}b. The cycle is compatible with a superconducting state.

The cycle, though, does not start at the zero of magnetization, which according to Bean's model \cite{Bean}, implies that a large number of vortices have entered the sample in preceding measurements. As before doing this particular cycle the sample had been annealed by heating it to the maximum available temperature, i.e. 400K, we conclude that the superconducting transition temperature is higher than this value.

The evolution with temperature of the ferromagnetic cycles obtained by subtraction of the measured diamagnetic linear contribution of  bulk graphite for S2 is shown on Fig.~\ref{Sigma}c. The cycles diminish in amplitude, collapsing above 150K . It is clear that the superconducting bulge is still there when ferromagnetism disappears, occupying almost all the thickness of the cycle at 200K.

The saturation moment M$_s$, defined in Fig.~\ref{Sigma}, is extracted from each curve and plot as a function of temperature to analyze its evolution (Fig.~\ref{Sigma}d). It is fitted by the known \cite{Carley} phenomenological expression(\ref{Ferro}) for a ferromagnet below the Curie temperature T$_0$ , M$_0$ being the saturation magnetization at 0K. 
\begin{equation}
 {M_s(T,H)=M_0(H)[1-(T/T_0)^2]^{1/2}  }
 \label{Ferro}
 \end{equation} 
The temperature dependence for the three samples is similar but not identical. For sample S1 T$_0$=230K, S2 T$_0$=150K, and even if it does not disappear completely for S3 T$_0$=350K. The values of the saturation magnetization at 0K are: for S1 M$_0$=0.0017emu/gr(3.4x10$^{-5}$$\mu_B$ per carbon atom), for S2  M$_0$=0.009 emu/gr (1.8x10$^{-5}$$\mu_B$ per carbon atom) and S3 M$_0$=0.01emu/gr(2x10$^{-5}$$\mu_B$ per carbon atom). These values correspond in the average to  $\sim$1x10$^{-5}$ $\mu_B$ per carbon atom.

\begin{figure}[ht]
\centering
\includegraphics[width=17cm]{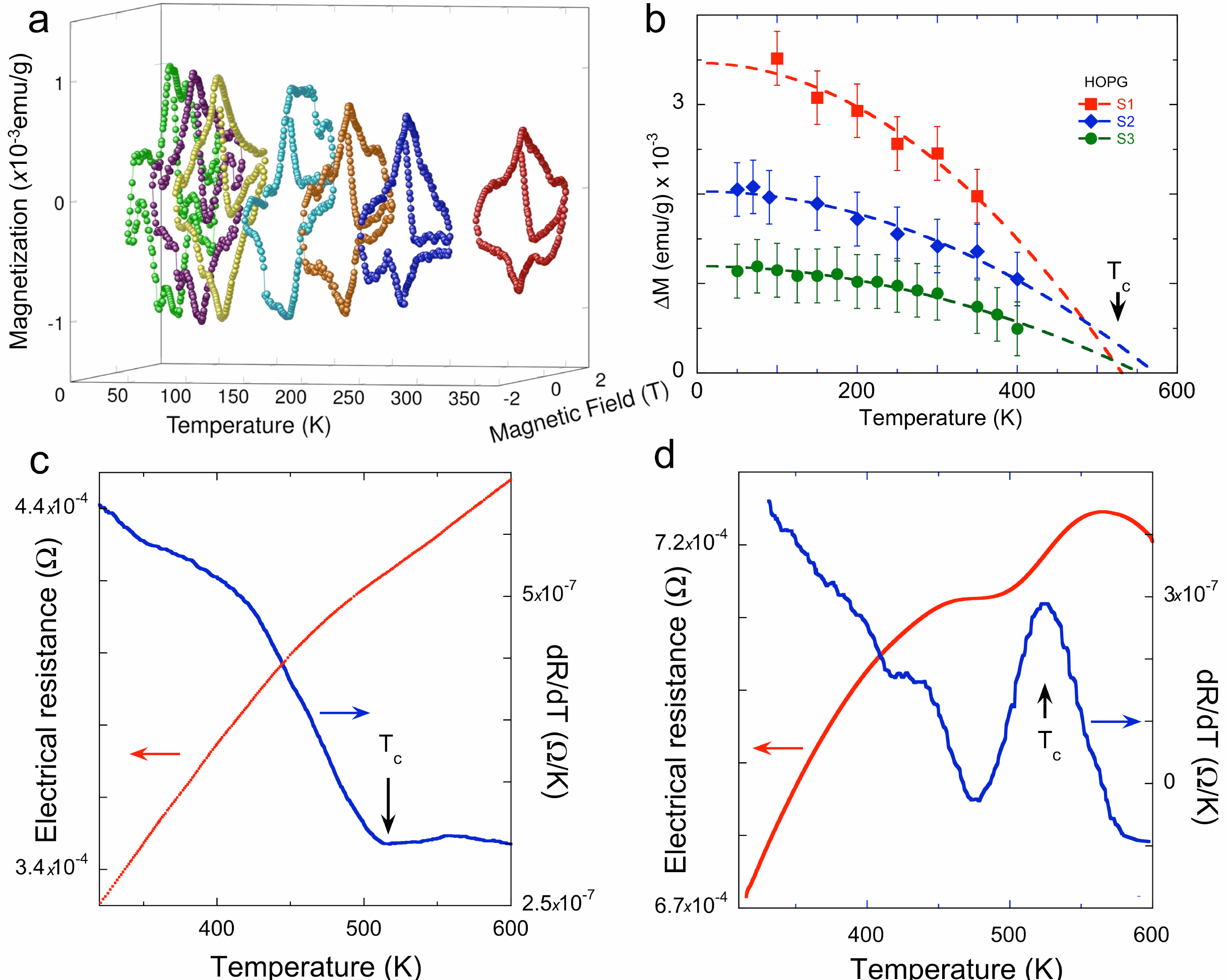}

\caption{(a) 3D plot of the superconducting-like hysteresis loops obtained by subtraction of the diamagnetic graphite contribution and the ferromagnetic sigmoidal fit for S2. We observe that the cycles decrease in amplitude with increasing temperature. (b) Temperature dependences of the amplitude $\Delta$M of the cycles at H=-0.1T, as defined in Fig.~\ref{Sigma}, for S1 (red squares), S2 (Blue diamonds) and S3 (green circles). As $\Delta$M(H,T)$\sim$J$_c$(H,T) we fit the dependences with the known expression (\ref{Jc}) for the temperature dependence of  J$_c$(H,T) (dashed curves), that allows determining an average T$_c$ $\sim$ 550$\pm$50K for these HOPG samples.(c) (red line) Temperature dependence of the electrical resistance with four aligned contacts in sample S4; (blue line) derivative of the electrical resistance as a function of temperature showing a transition with an onset T$_c$$\sim$521$\pm{50}$K).(d) (red line) Temperature dependence of the electrical resistance in van der Pauw contact configuration for S5 (see discussion on resistance measurements in Materials and Methods); (blue line) derivative of the electrical resistance as a function of temperature showing a transition at mid transition T$_c$$\sim$523$\pm{20}$K, and an onset T$_c$$\sim$560$\pm{30}$K.}
\label{DMRES} 
\end{figure}

The evolution with temperature of the  cycles obtained by subtraction of the bulk graphite linear diamagnetism and, only below 150K, of the ferromagnetic sigmoidal evolution is shown for sample S2 on Fig.~\ref{DMRES}a. It is clear that the amplitude of the cycle decreases with increasing temperature but it is still non-zero at 350K. Assuming Bean's model\cite{Bean}, then $\Delta$M(H,T), as defined in Fig.~\ref{Sigma}b, is proportional to J$_c$(H,T), that follows with temperature the expression (\ref{Jc}). We thus plot on Fig.~\ref{DMRES}b the $\Delta$M(H,T) as a function of temperature for samples S1, S2 and S3, and fit them with expression (\ref{Jc}). The result yields for three samples an average T$_c\sim$550$\pm$20K.
\begin{equation} \label{Jc}
 {J_c(T)=J_{c0}[1-(T/T_c)^ 2]}
  \end{equation}

With these T$_c$ values as a guide, we measured the electrical resistance at high temperatures on several samples.  The results on two pieces of the HOPG ZYA slab, S4 and S5, are shown on Fig.~\ref{DMRES}c and d, respectively. A clear kink in the linear variation of the resistivity of S4 and a small jump for S5 is observed. Calculating the derivative allows a better precision in the determination of its position. Its shape point to an onset T$_c$=521$\pm$50K for S4 and 560$\pm$30K for S5 and a mid T$_c$=523$\pm$20K for S5. These values coincide within experimental error with the T$_c\sim$550K determined previously. Cycling above 700K cause the disappearance of the anomaly in all cases.

Even though there is a kink or a small jump in the resistance, there is no percolation. Considering that superconductivity is granular, percolation should be indeed difficult, if not impossible, to obtain. As discussed previously, superconductivity may be due to ABC/ABA stacking faults \cite{Hentrich} or  "magic angle" between different graphite monocrystals in the HOPG matrix\cite{Heikila,Guinea}.

We now describe results on the dirtiest and more disordered samples of our lot, expanded carbon foils used as gaskets in refineries or car engines. Grafoil${\textregistered}$ \cite{Shane}and Papyex${\textregistered}$ are similar materials, both being manufactured in a process whereby finely ground particles of mined graphite are intercalated with several substances, then rinsed with water to produce a residue compound. This compound is then subjected to intense heat (900-1200 $^{\circ}$C) to initiate exfoliation. The exfoliated graphite is then compressed via rollers into a flexible foil-like material which is able to hold together without the aid of binding agents. Further heating (600$^{\circ}$C) is applied to drive off impurities and any remnant intercalant. It is another intercalation-deintercalation process, similar in principle to the one described in Ref.~\onlinecite{Layek}, but more violent.

\begin{figure}[ht]
\centering
\includegraphics[width=\linewidth]{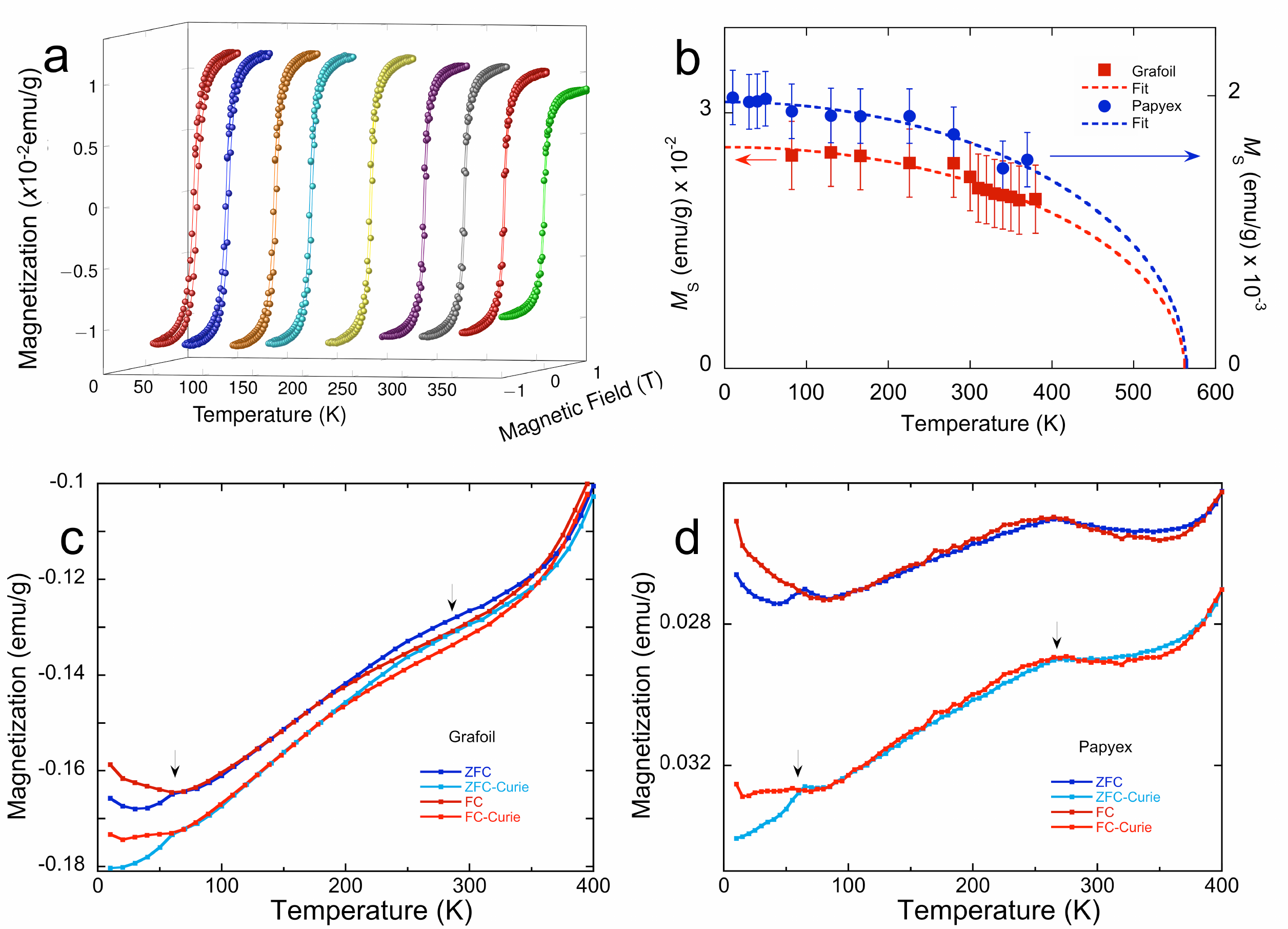}
\caption{ (a) 3D plot of the ferromagnetic cycles  of the Grafoil sample S6, obtained by subtracting only the linear diamagnetic graphite contribution from the raw data. We observe that the cycles decrease in amplitude with increasing temperature. (b) Saturation magnetizations as a function of temperature for samples Grafoil S6 (red squares) and Papyex S7 (blue circles) . Dashed curves: fits with the phenomenological function (\ref{Ferro}).(c) Zero field cooled and field cooled temperature magnetization measurements at 1T for Grafoil sample S6. We observe two transitions at $\sim$270K and $\sim$70K. Upper curves: For the raw data we observe at low temperatures a Curie paramagnetic law that can be fitted and subtracted. Lower curves: upon subtraction of the Curie law we obtain what seems to be the signature of a superconducting transition at $\sim$70K.(d) Similar as (c) but at 3T for Papyex sample S8. }
\label{FerroM} 
\end{figure}

Zero field cooled (ZFC) and field cooled (FC) measurements on Grafoil and Papyex samples are shown on Fig.~\ref{FerroM}a and b. Two anomalies at around 70K and 260K are present in both samples. Also the diamagnetic signal continues to increase at the highest measured temperature, suggesting another transition at higher temperatures. In both samples, at low temperatures we observe a Curie law type increase, which can be fitted and subtracted from the data. This behavior can be attributed to magnetic impurities, known to exist\cite{Quique} in such industrial products. In both cases the impurities' Curie temperature T$_{0{_{imp}}}$$\sim$ - 20$\pm$5K, indicating weak antiferromagnetic interactions among impurities.
Upon subtraction, the anomaly at low temperatures shows a behavior that has strong similarities with what could be expected from measurements on a superconductor with such a T$_c$. The transition at higher temperatures is less field dependent. 

The evolution with temperature of the ferromagnetic cycles of Grafoil samples obtained by subtraction of the measured diamagnetic linear contribution of bulk graphite is shown for S6 on Fig.~\ref{FerroM}c. The amplitude of the cycles slowly diminish with temperature. Similar measurements done on Papyex samples yield an analog behavior. We plot on Fig.~\ref{FerroM}d the temperature dependence of the magnetization saturation at 1T for two typical Grafoil and Papyex samples. By fitting with expression (\ref{Ferro}) both dependences we obtain similar values of T$_0$=562$\pm$50K. The values of the saturation magnetization at 0K are: for S6 M$_0$=0.026emu/gr(5.2x10$^{-5}$$\mu_B$ per carbon atom) and for S8  M$_0$=0.002 emu/gr (0.4x10$^{-5}$$\mu_B$ per carbon atom).

\begin{figure}[ht]
\centering
\includegraphics[width=\linewidth]{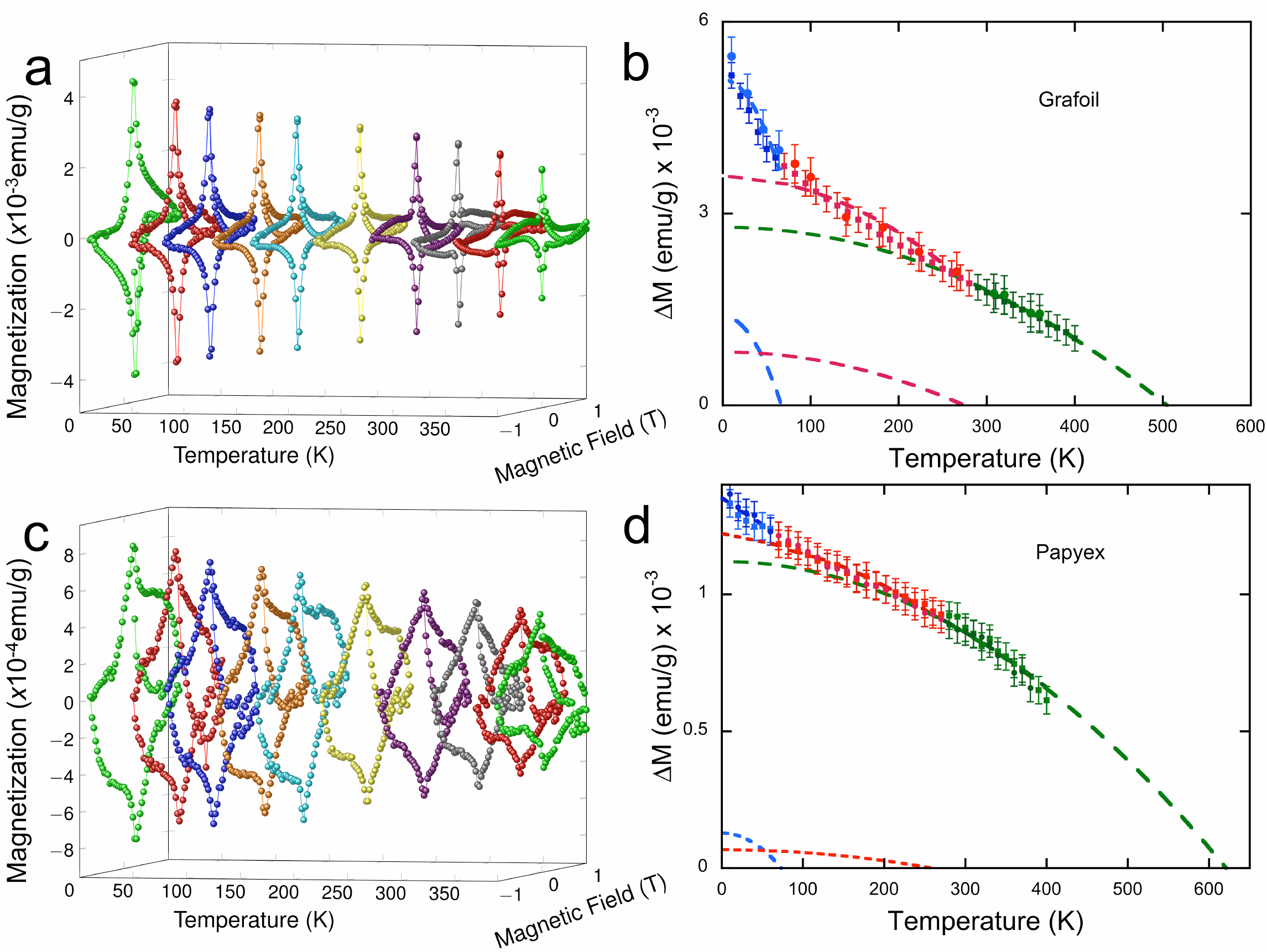}
\caption{ (b) Evolution of the superconducting-like hysteresis loops extracted by subtraction of the linear graphite contribution and a sigmoidal contribution (as in Fig.~\ref{Sigma}) for Grafoil sample S9. (b) Temperature dependences of the amplitude $\Delta$M of the cycles at H=-0.08T, as defined in Fig.~\ref{Sigma}, for Grafoil samples S6 and S9. As $\Delta$M(H,T)$\sim$J$_c$(H,T) we fit the dependences with the known expression (\ref{Jc}) for the temperature dependence of  J$_c$(H,T) (dashed curves), that allows determining three average  T$_c$ $\sim$ 510$\pm$50K, T$_c$$\sim$275K$\pm$15 and T$_c$ $\sim$69$\pm$5K (details in text). (c) Evolution of the superconducting-like hysteresis loops extracted by subtraction of the linear graphite contribution and a sigmoidal contribution (as in Fig.~\ref{Sigma}) for Papyex sample S8. (d) Temperature dependences of the amplitude $\Delta$M of the cycles at H=-0.08T, as defined in Fig.~\ref{Sigma}, for Papyex samples S8 and S10. As $\Delta$M(H,T)$\sim$J$_c$(H,T) we fit the dependences with the known expression (\ref{Jc}) for the temperature dependence of  J$_c$(H,T) (dashed curves), that allows determining three average  T$_c$ $\sim$ 623$\pm$50K, T$_c$$\sim$257K$\pm$15 and T$_c$ $\sim$77$\pm$5K (details in text).}
\label{PapGraf} 
\end{figure}

The temperature dependence of the superconducting-like hysteresis loops obtained by subtraction of the bulk graphite linear diamagnetism and of the ferromagnetic sigmoidal evolution is shown for the Grafoil sample S8 on Fig.~\ref{PapGraf}a and for Papyex sample S11 on Fig.~\ref{PapGraf}c. We then plot the evolution of $\Delta$M(H,T), as defined in Fig.~\ref{Sigma}b, as a function of temperature for the Grafoil and the Papyex samples on  Fig.~\ref{PapGraf}c and  on Fig.~\ref{PapGraf}d. The overall shape is sharply different from those of HOPG samples. The reason is that, as temperature decreases, we have the transitions at 260K and 70K that appear and start contributing to the thickness of the superconducting-like hysteresis loop that we measured. To separate the effect on the cycles for the different transitions, we first fit with expression (\ref{Jc}) only the points above 260K. This fit corresponds to the evolution of the highest temperature transition. Next, we fit the points between the two low temperature transitions, 260K and 70K, and fit with expression (\ref{Jc}) plus the fit obtained for the high temperature transition. Finally, we choose the points below 70K and fit with expression (\ref{Jc}) plus the fit obtained for the high temperature transition, plus the fit obtained for the intermediate region.

We show the result of the fits on for Grafoil samples S6 and S9 Fig.~\ref{PapGraf}c and for Papyex samples S7 and S10 on Fig.~\ref{PapGraf}d. The green dashed curves correspond to the high temperature transition, $\sim$510$\pm$50K for the Grafoil samples and $\sim$ 623$\pm$50K for the Papyex samples, the red dashed curves to the intermediate transition $\sim$275K$\pm$15 and $\sim$257K$\pm$15, respectively,  and the blue dashed lines to the low temperature transition, $\sim$69$\pm$5K and $\sim$77$\pm$5K, respectively. We show them added (on the experimental points) and separated (at the bottom of the graph). 

\textbf{Discussion}

It is interesting to compare the low and intermediate temperature results with those reported for IDI samples in Ref.~\onlinecite{Layek}. There, we have also two transitions at temperatures near to the ones observed here, namely 110K and 245K. Considering that the synthesis methods are not fundamentally different we would expect similar transition temperatures.  While the low T$_c$ for Grafoil and Papyex is more than 30\% smaller than the IDI value, the intermediate T$_c$ is less than 10\% higher. It can be that in IDI samples the residual potassium induces a doping different to the one operating in gasket foil samples. On the other hand, perhaps the difference is an effect of the magnetic impurities responsible for the Curie law, absent in IDI samples, lowering only the low T$_c$ value. 

We observe that for  the HOPG, Grafoil and Papyex samples the high temperature transition is, within the error of the determination, the same at around 550K. It is not astounding that hidden granular superconductivity has been so difficult to detect having such large T$_c$ values compared to room temperature, the highest temperature available in many laboratories that study superconductivity. 
Furthermore, electrical resistance measurements showed that the observed anomaly disappears by cycling above 700K. As our measurements in the optical oven are very fast (10-20K per minute), it is probable that the crystallites responsible for the anomaly, and superconductivity, are modified at these temperatures. Most probably this is due to changes in doping, as annealing of graphite defects can only be attained nearer to the graphitizing temperature. This fact will certainly be an obstacle for detailed studies around T$_c$.

We must note that the observed ferromagnetism possibly cannot be totally attributed to ferromagnetic impurities. Firstly, detailed analysis\cite{EsquiFerro1,EsquiFerro2} have determined on different types of HOPG samples, including those made by the same company as the ones measured here, that impurities would give a lower measured value. Secondly, and more important, it has been shown that annealing HOPG to temperatures near to the graphitizing temperature, 2400$^{\circ}$C, extinguishes\cite{Hebard} most of the ferromagnetic signal, implying an origin due to graphite defects. On the other hand,  the presence of the magnetic impurities (determined through a Curie law) boosts T$_0$ to a value that, within measuring error, is the same as the high temperature T$_c$. This fact is intriguing. It may be just due to some magnetic impurity. If we hypothesize that it is intrinsic, it might tell us that the maximum value of the Curie temperature of ferromagnetic state issued from the same interaction as the superconductivity that we observe is identical to the maximum superconducting transition temperature T$_{0_{max}}$=T$_{c_{max}}$.  Finally, within the assumption of an intrinsic carbon ferromagnetism, as the number of carriers for graphite is 2.4x10$^{-5}$ electrons and 1.8x10$^{-5}$ holes per carbon atom\cite{McClure}, it seems that a sizable number of carriers are spin polarized, although it is clear that the exact number depends on the particular geometry of each sample.

The origin of the observed granular superconductivity is still blurry and speculative. The spread of T$_c$'s above room temperature can obviously be due to the fact that they are obtained from the extrapolation of a phenomenological fit. From the electrical resistance measurements, though, there seems to be differences due to, perhaps, doping. For the low temperature transitions at around 100K and 250K, seen only in samples obtained from the IDI process that renders bulk twisted graphite\cite{Layek} and gasket foils made from compressed expanded graphite, the origin may be related to Moiré type defects. While in pure and well crystallized graphite, omnipresent rhombohedral stacking faults have been held as responsible\cite{Heikila}, even though Moiré defects also exist in well crystallized graphite\cite{Boi}. Nonetheless, the physics behind seems to be in both cases that of flat bands. As flat band superconductivity is very strong coupling \cite{Volovik}, the expression for the superconducting gap instead of being {$\Delta \sim$exp[-1/V$\rho$(0)]} is  $\Delta \sim$ V$\rho$(0), where V is the superconducting interaction  and $\rho$(0) the density of states at the Fermi level,  proportional to the flatness of the band. Thus, while for normal coupling superconductivity an increase of $\rho$(0) is strongly attenuated by the exponential, in flat band superconductivity it is not, and T$_c$$\sim\Delta$ is not bounded. Our measurements might be an example. In any case much work is still necessary to confirm, pin down and understand the origin of the superconductivity (and ferromagnetism) in graphite.

\textbf{Materials and Methods}

HOPG samples were obtained from portions of ZYA grade HOPG from Union Carbide derived from a neutron monochromator and of ZYH grade from Neyco.  All the HOPG samples were measured with the field parallel to the \textbf{c} graphitic axis. de Haas-van Alphen oscillations of S2 show a clear frequency at 4.8T corresponding to the hole’s Fermi surface of 4.7x1012 cm$^{-2}$ of Bernal graphite, a mobility of 2x10$^4$ cm$^2$/Vs at 5K can be estimated from onset magnetic field of the quantum oscillations. These results confirm the high quality and homogeneity of the samples.

Grafoil${\textregistered}$  samples were cut from a large roll obtained from Union Carbide (presently NeoGraf), while the Papyex${\textregistered}$-N samples were extracted from a piece furnished by the Carbone Lorraine (presently Mersen) research laboratory. Both materials were thoroughly studied by neutron diffraction previous to studies of $^3$He adsorption on graphite\cite{Quique}. Contrary to HOPG ZYA and ZYH that have a mosaic spread of 0.4$^\circ$ and 3.5$^\circ$, respectively, in Grafoil and Papyex the spread is 30$^\circ$. For the magnetization measurements, stripes were cut that were then folded into a cuboidal shape.

Magnetization measurements were performed in a Quantum Design MPMS3,VSM-Squid Magnetometer, at IN, and two Quantum Design MPMS XL QD at IN and LPS. The sample holder were the thin plastic straws furnished by Quantum design, with no other addenda. 

High temperature resistivity measurements were performed in home-made fast optical furnaces at IN, in a four lead DC measurement with four tungsten elastic fingers to ensure the electric contact. Two different apparatus were used, one with four fixed  aligned contacts and the other with four fixed contacts in a van der Pauw configuration. For this last one only one configuration opposing current and voltage contacts was measured. This non-linear configuration can be more sensitive to high conductivity defects in a large surface, but the value or even the shape can be approximative if the alternative opposed current/voltage configuration is not measured at the same time, which was not possible in our apparatus. While the linear configuration can give better results only if they are correctly placed on the defect. Dozens of contact positioning had to be made to locate half a dozen "sweet spots" where transitions were observable.


\textbf{Acknowledgments}

MNR thanks  K. Hasselbach, C. Paulsen for discussions and  P. Monceau, J.E. Lorenzo-D\'iaz, M-A. M\'easson, O. Buisson, F. Levy-Bertrand for a critical reading of the manuscript. We thank H. Godfrin and M. d'Astuto for providing us with the Grafoil and Papyex samples and the ZYA monochromator, respectively, and A. Hadj-Azzem and J. Balay for technical help. SL acknowledges support from the French National Research Agency through the projects IRONMAN ANR-18-CE30-0018 and MNR from PRESTO ANR-19-CE09-0027.

\appendix

\bibliographystyle{aipnum4-1}

\newpage

\begin{table*}[ht]
	\begin{center}
		\caption{Sample description. }
		\label{tab:table1}
		\begin{tabular}{l|c|c|c|c|r} 
			\textbf{Sample}&\textbf{Type} & \textbf{Company} & \textbf{Measurement}&\textbf{T$_c$'s(K)}\\
			\textbf{number}&&&   &&\ \\
			\hline
			\textbf{S1}  &HOPG ZYA & Union Carbide & Magnetization&531K\\
			\textbf{S2} &HOPG ZYA & Union Carbide& Magnetization&573K\\
			\textbf{S3} &HOPG ZYH & Neyco& Magnetization&554K\\ 
			\textbf{S4} &HOPG ZYA & Union Carbide& Resistance&521K\\ 
			\textbf{S5} &HOPG ZYA & Union Carbide & Resistance&560K\\ 
			\textbf{S6}  &FGG Grafoil & Union Carbide & Magnetization&69K; 275K; 510K\\
			\textbf{S7} &FGG Papyex& Carbone Lorraine& Magnetization&67K; 266K; 534K\\ 
			\textbf{S8} &FGG Papyex & Carbone Lorraine& Magnetization&77K; 257K; 623K\\ 
			\textbf{S9}  &FGG Grafoil & Union Carbide & Magnetization&69K; 275K; 510K\\
			\textbf{S10} &FGG Papyex & Carbone Lorraine& Magnetization&77K; 257K; 623K\\ 
		\end{tabular}
	\end{center}
\end{table*}

\newpage
\textbf{Figures}
\

\end{document}